\documentstyle[aps,prb]{revtex}
\begin{document}
\draft

\title{On the Stability and Single-Particle Properties of
Bosonized Fermi Liquids}
\author{A. Houghton, H.-J. Kwon, and J. B. Marston}
\address{Department of Physics, Brown University, Providence, RI 02912-1843}
\date{October 18, 1993}
\maketitle

\begin{abstract}
We study the stability and single-particle properties of Fermi liquids in
spatial dimensions greater
than one via bosonization.  For smooth non-singular Fermi liquid
interactions we obtain Shankar's renormalization-group flows and
reproduce well known results for quasi-particle lifetimes.  We demonstrate
by explicit calculation that spin-charge separation does not occur when the
Fermi liquid interactions are regular.
We also explore the relationship between quantized bosonic excitations and
zero sound modes and present a concise derivation of both the spin and
the charge collective mode equations.
Finally we discuss some aspects of singular Fermi liquid interactions.

\end{abstract}

\section{Introduction}
Landau's Fermi liquid theory is an early example of what we would now call
bosonization.  The anticommuting operators which appear in the bare Hamiltonian
describing the interactions among fermions disappear in Landau's
effective theory.  Instead only c-number quasi-particle occupancies
appear in the semi-classical energy functional.  That the low energy
semi-classical behavior of the Fermi liquid can be described in terms of
these commuting variables suggests that a fully quantum bosonic
description is obtainable.

Indeed, the Fermi liquid state itself is an example of a zero
temperature quantum critical fixed point\cite{PWAbook}.
This fixed point is characterized by infinite U(1) symmetry which is
{\it not} exhibited by the bare Hamiltonian.  The infinite U(1) symmetry simply
reflects the conservation of quasiparticle occupancy at each point on the
Fermi surface\cite{Haldane,Tony}.
Shankar\cite{Shankar} has used the functional
renormalization group (RG) approach
to show that the Fermi liquid state
is a generic feature of interacting fermions, at least at weak coupling
and in the absence of the usual
superconducting and charge- and spin-density wave instabilities.
By establishing
rigorous bounds, other workers
have studied the stability question at all orders in perturbation
theory, but under more restrictive conditions such as a perfectly circular
Fermi surface\cite{Others}.

Haldane has asserted recently that a fully quantum description of
Fermi liquids in dimensions greater than one is obtainable
via bosonization\cite{Haldane}.  This viewpoint has been elaborated on
by two of us\cite{Tony}.  In the present paper we continue to develop this
theory first by showing that Shankar's renormalization group result is
obtained easily in the bosonized picture.  Next we investigate the bosonic
excitations in more detail.
We show that collective modes are obtained in a semi-classical
limit; furthermore the calculation of the single-particle boson Green's
function yields information about
the quasiparticle properties.  In particular, by using the bosonization
transformation to determine the fermion quasiparticle
propagator we obtain well-known results for the fermion
self energy: the imaginary
part is proportional to $\omega^2 \ln |\omega|$ in two dimensions and just
$\omega^2$ in three dimensions.  We emphasize that the bosonization method
yields non-perturbative information, so a natural next step would be to
use it to study the effects of singular interactions.
We comment on the nature of two such singular interactions.

\section{Renormalization Group Analysis in the Bosonic Basis}

Now that we know how to
bosonize Fermi liquids we may use this picture to investigate
the stability of the zero-temperature Fermi liquid fixed point
to perturbations.
First we reproduce the renormalization group results of
Shankar\cite{Shankar} in the bosonic basis.
Three channels of fermion two-body interactions are marginal in the RG sense:
forward scattering zero-sound (ZS), exchange scattering (ZS$^\prime$),
and Cooper pairing (BCS).
For simplicity we consider a system of spinless fermions in two dimensions and
a circular Fermi surface.  The second assumption eliminates
the possibility of nesting instabilities in the zero-sound channels which might
produce charge or spin density waves.  We also assume that the
BCS coupling function $V_{BCS}(S)$ is rotationally invariant.
The BCS interaction pairs particles of equal but opposite
momenta.  For now we turn off the two zero sound channels; later we show
that these channels have no effect on the renormalization of the BCS
interactions.

Fermi fields $\psi$ may be expressed\cite{Tony} in terms of the boson fields
$\phi$ as:
\begin{equation}
\psi(S;{\bf x}) = {1\over\sqrt{V}}~ \sqrt{{\Omega}\over{a}}
e^{i{\bf k}_S{\bf \cdot x}}
 \exp \{i{\sqrt{4\pi}\over \Omega} \phi(S;{\bf x})\}~ O(S),
\label{bosonization}
\end{equation}
where the dependence on time, $t$, is included implicitly in
the spatial coordinates ${\bf x}$.  $S$ runs from 0 to $2 \pi$ and
labels the patch
on the Fermi surface with momentum ${\bf k}_S$.  $V$ is the volume of the
system which equals $L^2$ in two dimensions; the factor of $V^{-1/2}$
is introduced to keep the
fermion anticommutation relations canonical.
Both the $\psi$ and $\phi$ fields live inside a squat
box centered on $S$ with height $\lambda$ in the radial (energy) direction
and width
$\Lambda$ along the Fermi circle.  These two scales must be small in the
following sense: $k_F >> \Lambda >> \lambda$.  We satisfy these limits
by setting $\lambda \equiv k_F/N$ and $\Lambda \equiv k_F/N^\alpha$ where
$0 < \alpha < 1$ and $N \rightarrow \infty$.
The quantity $a$ in the bosonization formula Eq. (\ref{bosonization})
is a real-space cutoff given by $a \equiv 1/\lambda$.  Here
$\Omega \equiv \Lambda (L/2 \pi)^2$ equals the number of states in the squat
box divided by $\lambda$.  Finally, $O(S)$ is an ordering operator
introduced\cite{Tony,Luther}
to maintain Fermi statistics in the angular direction along
the Fermi surface.  (Anticommuting statistics are obeyed automatically
in the radial direction.)

With this connection between the fermion and boson fields
we may check a number of relationships.  For example, the fermion fields
obey canonical anticommutation relations:
\begin{equation}
\{\psi(S; {\bf x})~ ,~ \psi^\dagger(T; {\bf y}) \} = \delta_{S,T}~
\delta^2({\bf x - y})
\end{equation}
because the boson fields in configuration space obey commutation
relations:
\begin{eqnarray}
[\phi(S; {\bf x})~ ,~ \phi(T; {\bf y})]
&=& {i\over4}~ \Omega^2~
\delta_{S,T}~ \epsilon(\hat{\bf n}_S \cdot [{\bf x - y}])~ ;\
|x_\perp - y_\perp| << 1/\Lambda
\nonumber \\
&=& 0~ ;\ |x_\perp - y_\perp| >> 1/\Lambda\ .
\label{comm}
\end{eqnarray}
Here $\perp$ denotes directions perpendicular to the surface normal
$\hat{\bf n}_S$ at patch $S$, and
$\epsilon(x) = 1$ for $x > 1$; otherwise it equals -1.
Normal ordered charge currents are defined in configuration space
in terms of the both the Fermi and
Bose fields as:
\begin{eqnarray}
J(S; {\bf x}) &=& V :\psi^\dagger(S; {\bf x}) \psi(S; {\bf x}): \nonumber \\
&\equiv& V \lim_{\epsilon \rightarrow 0}~
\{ \psi^\dagger(S; {\bf x} + \epsilon \hat{{\bf n}}_S)~ \psi(S; {\bf x})
- \langle \psi^\dagger(S; {\bf x} + \epsilon \hat{{\bf n}}_S)~ \psi(S; {\bf x})
\rangle \} \nonumber \\
&=& \sqrt{4 \pi}~ \hat{{\bf n}}_S {\cdot \nabla} \phi(S; {\bf x})\ .
\label{curx}
\end{eqnarray}
The momentum-space charge current is defined by:
\begin{equation}
J(S; {\bf q}) \equiv \sum_{\bf k} \theta(S; {\bf k + q})~
\theta(S; {\bf k})~ \{ \psi^{\dagger \alpha}_{\bf k + q}~
\psi_{\alpha {\bf k}} - \delta^3_{\bf q, 0}~ n_{\bf k} \}
\label{curk}
\end{equation}
where $\theta(S; {\bf k}) = 1$ if ${\bf k}$ lies inside the squat box
of dimensions $\lambda \times \Lambda$ centered at $S$ and equals zero
otherwise.  Given
this definition, plus the fact that the Fermi fields in momentum and real
space are related in the usual way to preserve the canonical anticommutation
relations with conventional normalization,
\begin{equation}
\psi(S; {\bf x}) = {{1}\over{\sqrt{V}}}~ \sum_{\bf k}~ \theta(S; {\bf k})~
e^{i {\bf k \cdot x}}~ \psi({\bf k})\ ,
\end{equation}
the two currents are related by a Fourier transform:
\begin{equation}
J(S; {\bf x}) = \sum_{\bf q}~ e^{i {\bf q \cdot x}}~ J(S; {\bf q})\ .
\end{equation}
Both currents Eq. (\ref{curx}) and Eq. (\ref{curk})
are dimensionless.
The free Hamiltonian, written in terms of the Fermi fields,
may also be bosonized using
Eq. (\ref{bosonization}) and the result is quadratic in the $\phi $ fields:
\begin{eqnarray}
H_0  &=& v_F~ \sum_S \int d^2x~ \psi^{\dag}(S;{\bf x})~ \bigg{\{}{{\hat{\bf
n}_S
{\bf \cdot \nabla}}\over{\rm i}} - k_F\bigg{\}}~ \psi(S;{\bf x}) \nonumber \\
&=& {{4\pi~ v_F}\over{\Omega~ V}}~
\sum_S \int d^2x~ \{(\hat{\bf n}_S{\bf \cdot \nabla})~ \phi(S;{\bf x})\}^2 ~.
\end{eqnarray}
The unusual prefactor of ${4\pi~ v_F \over \Omega~ V}$
appearing in the bosonic Hamiltonian
compensates for the anomalous right hand side of the boson commutation
relations, Eq. (\ref{comm}), and thereby reproduces the correct spectrum.

The next step is to bosonize the BCS interaction.  To simplify the following
algebra we set the Fermi velocity equal to one ($v_F = 1$).
A fermion in patch $S$ of the Fermi surface
is paired with a fermion in patch $-S$ which is directly opposite patch
$S$.  (Note that in angular coordinates,
patches $S$ and $-S$ correspond to $\theta$ and $\theta + \pi$, {\it not}
$\theta$ and $-\theta$.)  The BCS action expressed in
terms of the four Fermi fields is:
\begin{equation}
 S_{int}[\psi, \psi^\dagger]~ =~
\sum_{  S ,T } \int dt~ d^2 x ~{{V_{BCS}(S-T)}\over{k_F}}~
\psi ^{\dag }
(-S;{\bf x})~ \psi ^{\dag}(S;{\bf x})~\psi (T;{\bf x})~ \psi (-T;{\bf x}) ~.
\end{equation}
Here, $S$ and $T$ only range over half of the Fermi surface to avoid double
counting the pair interactions. The dimensionless
coupling function $V_{BCS}$ must change sign under inversion because the
fermions are spinless (Pauli exclusion principle) so
$V_{BCS}(\theta)=-V_{BCS}(\theta +\pi)$; also the interaction must
be Hermitian so $V_{BCS}(S-T)
= V_{BCS}(T-S)$.  To avoid sign mistakes
it is important to keep track of the order of the
fermion operators during the transformation to bosons.
Formally the correct sign is set via the ordering operator $O(S)$ but in
practice it is easier to determine the sign by direct inspection of the
fermion operators.
To bosonize the interaction, each Fermi field is replaced by the right hand
side
of Eq. (\ref{bosonization}) which converts the interaction into the exponential
of four $\phi $ fields.
\begin{equation}
S_{int}[\phi]~ =~ ({{\Lambda}\over{2 \pi}})^2~ \sum_{ S , T } \int dt~ d^2x~
{{V_{BCS}(S-T)}\over{k_F~ a^2}}
\cos\bigg{[}
{\sqrt{4\pi } \over \Omega }(\phi _S ({\bf x })-\phi _T ({\bf x}))\bigg{]}~,
\end{equation}
where $\phi _S ({\bf x }) \equiv \phi (S;{\bf x}) + \phi(-S;{\bf x}).$

Before implementing the RG transformation, first we discuss scaling
at the zero loop level.
Since we are concerned with scaling in the direction parallel to the surface
normal $\hat{\bf n}_S$,
the integral over $\bf x$ space should be factorized into
separate integrals over
directions perpendicular and parallel to the Fermi surface normal.  So
$d^2x = dx_{\perp}~dx_{\parallel}$ with only $dx_{\parallel}$ changing under
scale transformations.  Thus: $\Lambda \rightarrow \Lambda$, $\lambda
\rightarrow \lambda / s$ and $a \rightarrow s~ a$ with $s > 1$.  Clearly,
$dx_{\parallel} \rightarrow s~ dx_{\parallel}$ and $dt \rightarrow s~ dt$.
The boson field $\phi$ is invariant under the scale transformation as this
leaves both the quadratic part of the action, $S_0$, and the BCS interaction
invariant (marginal).

Now we perform the renormalization group transformation on the BCS interaction
to derive the flow equation.
The fast parts of the $\phi(T; {\bf k})$ fields are integrated out of the
functional integral.  To be precise, modes with momenta
$ \lambda /2s < |{\bf k\cdot \hat{n}}_T| < \lambda /2 $ will be eliminated.
In practice since it is easier to carry out the calculation in real space,
instead we integrate out fields over short {\it distance} scales
$2 a < {\bf x \cdot \hat{n}}_T < 2 s~ a$.
Next, we rescale space and time:
${\bf \hat{n}}_T {\bf \cdot x} \rightarrow
s~ {\bf \hat{n}_T {\bf \cdot x}}$ and $ t \rightarrow s~ t$.
After performing these two operations we obtain the new
BCS interaction coefficients $V_{BCS}(S-T; s)$ and we may repeat the
process.

The integration over the fast modes is accomplished via the usual functional
integral:
\begin{equation}
\exp( -S[\phi ;s]) = \prod _{T} \int _{2 a < |{\bf x\cdot \hat{n}}_T|
< 2 s a} {\cal D}\not\!\phi (T;{\bf x}) ~\exp(-S_0 -S_{int}) ~. \label{RG}
\end{equation}
Since only the fast modes are integrated out it is convenient to break the
boson fields into two parts, the slow modes $\phi ^\prime $ and the
fast modes
$ \not\!\phi $.  Now we may express the right hand side of Eq. (\ref{RG}) as
$\langle \exp(-S_{int}) \rangle $ where
the contraction is performed only over the fast $\not\!\phi $ fields.
The renormalized interaction is obtained by treating the interaction
perturbatively and expanding $\langle \exp(-S_{int}) \rangle$ in powers of
$V_{BCS}$.  The first non-trivial term arises at second order,
${1\over 2} \langle S_{int}^2 \rangle$,
and since the interaction is diagonal in real-space, it is readily
evaluated with the use of the real-space boson correlation function\cite{Tony}:
\begin{eqnarray}
\langle \not\!\phi (T;{\bf x})~\not\!\phi (T;{\bf 0})
-\not\!\phi^2(T; {\bf 0})
\rangle &\simeq& {\Omega ^2 \over 4\pi }
\ln ({i a \over {\bf x\cdot \hat{n}}_T + i \tau})~ ;\ |x_\perp \Lambda| << 1
\nonumber \\
&\rightarrow& -\infty~ ;\ |x_\perp \Lambda| >> 1 .
\label{gphi0}
\end{eqnarray}
Here and in Eq. (\ref{RG})
we have made a Wick rotation to imaginary time to avoid the poles
along the real-time axis.
The evaluation of the cosine-cosine correlation function that appears in
${1\over 2} \langle S_{int}^2 \rangle$ is carried out by decomposing
terms of the form
\begin{equation}
\cos\bigg{[}
{\sqrt{4\pi } \over \Omega }(\phi _S ({\bf x })-\phi _T ({\bf x}))\bigg{]}
\end{equation}
into exponentials involving the slow and fast fields,
and then using the identity:
\begin{equation}
e^A~ e^B~ =~ : e^{A + B} :~ {\rm exp}
\langle A B + {{1}\over{2}} (A^2 + B^2) \rangle\ .
\end{equation}
Since $V_{BCS}(S-T) = V_{BCS}(T-S)$ we obtain:
\begin{eqnarray}
{1\over 2} \langle S_{int}^2 \rangle
&=& ({{\Lambda}\over{2\pi}})^3~ {{1}\over{(k_F a)^2}}~
\int d\tau~ d^2x~ \sum_{R, S}
\cos\bigg{[}{{\sqrt{4\pi}}\over{\Omega}}~ (\phi^\prime _{R} ({\bf x})-
\phi^\prime _{S}({\bf x}))\bigg{]}
\sum _T V_{BCS}(R -T)~V_{BCS}(T-S) \nonumber \\
&\times& {{1}\over{\pi}}\int^{\infty}_{-\infty} d\nu \int^{2sa}_{2a}
du_{\parallel}~ {1\over {u_\parallel}^2 + {\nu}^2}\ +\
{\rm \{irrelevant~ operators\}}~,\label{correction}
\end{eqnarray}
where $u_\parallel$ is a spatial variable in the direction parallel
to the surface normal at patch $T$ and $\nu$ is an imaginary time variable.
The second step is to replace all the variables with the rescaled ones. This
procedure does not change Eq. (\ref{correction})
since $V_{BCS}(S-T)$ is marginal.  Thus the $\beta$-function is:
\begin{equation}
{dV_{BCS}(R - S; s)\over d\ln (s) } = -{\Omega \over V~ k_F}
\sum_T V_{BCS}(R-T; s)
{}~V_{BCS}(T-S; s) ~.
\end{equation}
The equation may be diagonalized by a Fourier transform over the interval
$[0, 2\pi)$:
$V_m \equiv \int {d\theta \over 2\pi}~
 e^{i m \theta}~V_{BCS}(\theta)~ .$  Note that only odd modes appear due
to the requirement $V_{BCS}(\theta) = -V_{BCS}(\theta + \pi)$.  Then:
\begin{equation}
{dV_m\over d\ln(s)} = -{1\over 2}~ V_m^2 ~,
\end{equation}
which agrees with Shankar's result in the fermion
basis at one-loop order\cite{Shankar}.
Clearly, a BCS instability exists if any of the channels are attractive
($V_m < 0$).  If all the channels are repulsive, the Fermi liquid fixed point
is stable.

We now turn on the other two marginal interactions,
forward and exchange scattering, and ask whether the Fermi liquid interactions
$f_c(S-T)$ which involve the
ZS and ZS$^\prime$ channels alter the RG flows.
\begin{eqnarray}
H_0 &\rightarrow& H_0 + {{1}\over{2}}~ \sum_{S,T} \int d^2x~ f_c(S-T)~
\psi^{\dag}(S;{\bf x})~\psi(S;{\bf x})~\psi ^{\dag}(T;{\bf x})~\psi(T;{\bf x})
\nonumber \\
&=& H_0 + {{2 \pi}\over{V^2}}~ \sum_{S,T} \int d^2x~ f_c(S-T)~
[\hat{\bf n}_S {\bf \cdot
\nabla} \phi(S;{\bf x})]~[\hat{\bf n}_T {\bf \cdot \nabla} \phi(T;{\bf x})]~.
\label{intbos}
\end{eqnarray}
Unlike the
BCS interaction, the forward and exchange interactions are quadratic in
the boson fields and therefore parameterize different Gaussian fixed points,
each with the infinite U(1) symmetry.  This symmetry is reflected in the
fact that the Hamiltonian Eq. (\ref{intbos}) is invariant under
changes in the phase of the fermions by different amounts in each patch:
$\psi(S; {\bf x}, t) \rightarrow e^{i \theta(S)}~ \psi(S; {\bf x}, t)$.
Here we see the advantage of the bosonic representation: the
Fermi liquid parameters are incorporated in a non-perturbative way into $H_0$.
We carry out the calculation of the modified bosonic correlation function
in the next section;
here we just note that these modifications are sub-leading corrections
to scaling that do not influence the leading RG flows of the BCS interaction.
For example, though bosons in different patches are now correlated, this
correlation is only of order ${\Lambda \over k_F}$ times that of correlations
within the same patch.  So the leading behavior
exhibited in Eq. (\ref{correction}) is unchanged.  Thus we have the remarkable
result that the leading-order
stability of the Fermi liquid fixed point against Cooper
pairing is unaffected by the existence of either small or large Fermi liquid
parameters.

Actually, there is a sub-leading order instability: the Kohn-Luttinger
effect\cite{Shankar}.  The bare $V_m$ due to, say, a short-range
repulsive interaction are all positive but tend rapidly to zero at
large-$m$.  The ZS and ZS$^\prime$ channels, on the other hand,
generate irrelevant contributions in the BCS channel (down by a positive power
of $\Lambda/k_F$) which renormalize the bare BCS interaction
and therefore can make some of the $V_m$ slightly negative (unstable)
at sufficiently large-m.  Because of the small size of these
coefficients, this effect is important
only at extremely low temperatures and therefore we expect that the
essential physics remains controlled by the Fermi liquid fixed point.

\section{Interacting Bosons in the Semiclassical Limit}

Now that we have seen
that the Fermi liquid fixed point is stable in the absence
of attractive BCS interactions,
we turn to the problem of diagonalizing the bosonic Hamiltonian that describes
the fixed point.
As we shall see, the problem is not as simple as it might seem at first
because the current operators behave as both creation and annihilation
operators.  So we begin with an approximate semiclassical solution that
bypasses this difficulty and yields the familiar collective mode equation.

The particle-hole excitations of Fermi liquids have a bosonic character.
Furthermore, the excitations are of either the charge or spin type: the
bosonized Hamiltonian may be written as $H = H_c + H_s$ to exhibit this
factorization into charge and spin sectors.
The charge sector in D-dimensions (the volume $V = L^D$ now)
is described by a Hamiltonian that is bilinear in the current
operators\cite{Tony} $J({\bf S; q})$:
\begin{equation}
H_c = {{1}\over{2}}~ \sum_{\bf S, T} \sum_{\bf q} V_c({\bf S ,T; q})~
J({\bf S ; -q})~ J({\bf T; q})~, \label{hamc}
\end{equation}
where $ \Omega \equiv ({\Lambda L\over 2\pi })^{D -1} ({L\over 2\pi})$.
The Fermi liquid interactions are
$f_c({\bf S, T}) \equiv F_c({\bf S, T})/N(0)$,
where the density of states at the Fermi surface, summed over both spin
species,
is given by
$N(0) = {{k_F}\over{\pi v_F}}$ for the case of the two dimensional Fermi gas.
These interactions are incorporated into
$V_c$ as matrix elements that couple currents in different patches:
\begin{equation}
V_c ({\bf S, T; q}) = {{1}\over{2}}~ \Omega ^{-1} v_F~ \delta^{D-1}_{\bf S, T}
+ {{1}\over{V}}~ f_c({\bf S - T})~ . \label{intc}
\end{equation}
Note that with this definition, and given the
relationship between the currents and the $\phi$ fields,
Eq. (\ref{curx}), the charge Hamiltonian $H_c$ of
Eq. (\ref{hamc}) agrees (up to a factor of 2 due to spin)
with the form we found in the previous section,
Eq. (\ref{intbos}).
The charge currents obey the U(1) Kac-Moody relations:
\begin{equation}
[J({\bf S; q})~ ,~ J({\bf T; p})] = 2~ \delta^{D-1}_{\bf S, T}~
\delta^D_{\bf q + p, 0}~
\Omega~ {\bf q \cdot \hat{n}_S}\ ; \label{kacc}
\end{equation}
this algebra can be derived either from Eq. (\ref{comm}) and Eq. (\ref{curx})
or directly from Eq. (\ref{curk}) with the use of the canonical anticommutation
relations for fermions\cite{Tony}.
The quadratic form of this Hamiltonian implies immediately that it describes
a fixed point invariant under the scale transformations $\lambda \rightarrow
\lambda/s$.
Similarly, the spin-sector is described by:
\begin{equation}
H_s = {{1}\over{2}}~ \sum_{\bf S,T} \sum_{\bf q} V_s({\bf S, T; q})~
{\bf J}({\bf S; -q})~ {\bf \cdot~ J}({\bf T; q})\ \label{hams}
\end{equation}
where the spin currents commute with the charge currents and
obey the more complicated SU(2) Kac-Moody algebra\cite{Tony}:
\begin{equation}
$$[J^a({\bf S; q})~ ,~ J^b({\bf T; p})]
= {{1}\over{2}}~ \delta^{D-1}_{\bf S, T}~
\delta^{a b} \Omega~ {\bf q \cdot \hat{n}_S}~ \delta^D_{\bf q+p, 0}
+ i~ \delta^{D-1}_{\bf S, T}~
\epsilon^{a b c}~ J^c({\bf S; q + p})\ \label{kacs}
\end{equation}
($\{a, b, c\} = \{x, y, z\}$ label the three components of the spin).
Fermi liquid spin-spin interactions $f_s$
appear in the Hamiltonian
as coefficients that couple spin currents in different patches:
\begin{equation}
V_s({\bf S, T; q}) = {{2}\over{3}}~
v_F({\bf S})~ \Omega^{-1}~ \delta^{D-1}_{\bf S, T}
+ {{1}\over{V}}~ f_s({\bf S, T})\ .
\label{twothirds}
\end{equation}
Here we consider only the case of Fermi liquid interactions which are
independent
of the wavevector $\bf q$.  For regular interactions, $\bf q$ dependence
only gives rise to additional irrelevant operators which do not change the
behavior of the system at leading order.
With singular interactions, on the other hand,
divergences arise as ${\bf q} \rightarrow 0$ and these divergences may
in some instances introduce relevant interactions that destroy Fermi liquid
behavior.

The equations of motion for the charge and spin currents yield the
corresponding collective mode equations in the semi-classical limit.
Using the Heisenberg equations of motion and the U(1) Kac-Moody algebra we
readily obtain:
\begin{eqnarray}
i {{\partial}\over{\partial t}} J({\bf S; q}) &=& [J({\bf S; q}),~ H_c]
\nonumber \\
&=& v_F~{\bf q \cdot \hat{n}_S}~ J({\bf S; q}) + {\bf q \cdot \hat{n}_S}~
{{2 \Omega}\over{V}}~ \sum_{\bf T}~ f_c({\bf S - T})~ J({\bf T; q})\ .
\label{motc}
\end{eqnarray}
for the charge sector.  The first term on the right hand side has its origin in
the free dispersion relation for particle-hole pairs of momentum $\bf q$
at patch $\bf S$.  The second term couples currents in different patches.
Note that
\begin{equation}
{{\Omega}\over{V}}~ \sum_{\bf T} = \int {{d^Dk}\over{(2 \pi)^D}}~
\delta(|{\bf k}| - k_F)
\end{equation}
so the second term reduces to the usual
integral over the Fermi surface in the $N \rightarrow \infty$ continuum limit.
The equation of motion for the non-abelian spin currents contains, in
addition to these two terms, a third term which makes the spins precess,
when the system is magnetically polarized,
in the local internal magnetic field\cite{Baym}:
\begin{eqnarray}
i {{\partial}\over{\partial t}} J^a({\bf S; q}) &=& [J^a({\bf S; q}),~ H_s]
\nonumber \\
&=& v_F~{\bf q \cdot \hat{n}_S}~ J^a({\bf S; q}) + {{1}\over{2}}~
{\bf q \cdot \hat{n}_S}~
{{\Omega}\over{V}}~ \sum_{\bf T}~ f_s({\bf S - T})~ J^a({\bf T; q})
\nonumber \\
&-& {{\rm i}\over{V}}~ \epsilon^{a b c}~ \sum_{\bf k}~ J^b({\bf S; k})~
\sum_{\bf T}~ f_s({\bf S - T})~ J^c({\bf T; q - k})\ .
\label{mots}
\end{eqnarray}
Note that the factor of $2/3$ multiplying the Fermi velocity $v_F$
in the free part of Eq. (\ref{twothirds})
does not appear in Eq. (\ref{mots}).  The origin of the $2/3$ factor is easy to
understand in the Sugawara construction of the free fermion Hamiltonian out
of current bilinears\cite{sugawara}: it reflects the
SU(2) invariance of the spin currents which permits the replacement
${\bf J(S; q) \cdot J(S; -q)} \rightarrow 3 J^z({\bf S; q}) J^z({\bf S; q})$
for the purpose of computing the spectrum.  The derivation
given here, on the
other hand, does not rely on this argument as spin rotational invariance
is respected explicitly.  Rather, the factor of $2/3$ in Eq. (\ref{twothirds})
cancels contributions to the free spectrum which arises from both the
$\delta^{a b}$ and the $i~ \epsilon^{a b c}$ terms in the non-Abelian anomaly.

When an external magnetic field is applied, and $|{\bf q}|$ is small,
the third term in Eq. (\ref{mots}) dominates.
In the opposite limit of zero applied magnetic
field, the spin equation is identical in form to the charge equation.
Evidently bosonization captures all of the physics of charge
and spin collective modes.  The derivation is straightforward and relies
only on the existence of the quadratic Hamiltonian and the
Abelian and non-Abelian Kac-Moody commutation relations.
In fact this approach may be useful in the study of
highly spin-polarized Fermi liquids, a problem which has
been examined recently by Meyerovich and Musaelin via the Green's function
approach\cite{Alex}.

As it stands these operator equations are exact, at least in the $N \rightarrow
\infty$ limit in which the Kac-Moody algebras Eq. (\ref{kacc}) and Eq.
(\ref{kacs}) become exact.
The difficulty in obtaining exact solutions to either of these equations
originates in the fact that neither algebra is equivalent to the canonical
commutation relations for harmonic oscillators.  Even the U(1) charge current
algebra, Eq. (\ref{kacc}), is non-trivial because the right hand side,
the ``anomaly,'' has indeterminate sign as $\bf q \cdot \hat{n}_S$ can be
either positive or negative.  Therefore, even to diagonalize Eq. (\ref{motc})
requires a generalized Bogoliubov-unitary transformation which appears
not to be reducible into a product of
separate Bogoliubov and unitary transformations.  This
difficulty is in contrast to that found in one spatial dimension
where a simple $2 \times 2$
Bogoliubov transformation is sufficient to decouple the currents
associated with the left and right Fermi points\cite{Tony}.
Of course the equation of motion for the spin currents, Eq. (\ref{mots}), is
even more difficult to solve as it is non-linear.

In the next section
we will diagonalize the charge current equation of motion by resumming the
perturbative expansion for the propagator.
First we find the semiclassical solution by
taking the expectation value of both sides of these equations and identifying
\begin{equation}
\langle J({\bf S; q}) \rangle \equiv u({\bf S; q}) \label{expc}
\end{equation}
and
\begin{equation}
\langle J^a({\bf S; q}) \rangle \equiv S^a({\bf S; q}) \label{exps}
\end{equation}
as the
amplitudes for charge and spin collective modes.  Note that the collective
mode amplitudes are real-valued in x-space because
$J^\dagger({\bf S; q}) = J({\bf S; -q})$.  The replacement of the current
operators by the c-numbers $u$ and $\bf S$ may be accomplished
formally by introducing a coherent state basis that spans the space of
volume-preserving geometric distortions of the Fermi surface\cite{Fradkin}.
For example, the charge collective mode coherent states
are generated by exponentials of the charge current operator:
\begin{equation}
| \Psi[u] \rangle = \exp \bigg{\{} \sum_{\bf S}~ \sum_{\bf q}~
\theta({\bf q \cdot \hat{n}_S})~ {{u({\bf S; q})}\over{2 \Omega {\bf q \cdot
\hat{n}_S}}}~ J({\bf S; -q}) \bigg{\}} |0\rangle
\label{coher}
\end{equation}
where $|0\rangle$ represents the quiescent Fermi liquid.  A simple computation
then shows that
\begin{equation}
{{\langle \Psi[u] | J({\bf T; p}) | \Psi[u] \rangle}\over
{\langle \Psi[u] | \Psi[u] \rangle}} = u({\bf T; p})
\end{equation}
consistent with our definition Eq. (\ref{expc}).
In the spin equation we also must decouple the expectation value of
the product of two spin current operators [the third term on the right hand
side of Eq. (\ref{mots})] into the product of expectation
values $\bf S({T; k}) \times
S({T; q - k})$.  This decoupling is exact in the
semiclassical limit of a macroscopically occupied zero sound spin mode.

To shed light on the relationship between the semiclassical limit and the
quantum regime, we present an alternative derivation of the collective
mode equation based on the usual identification of the pole in the
two-point Green's function, but now
in the presence of a background collective mode field
$u({\bf S; q})$.  For simplicity we focus on the charge sector.  First
note that $H_c$ is diagonal in $\bf q$ space: the Hilbert space breaks up
into a direct product of subspaces with different $\bf q$ and  $\bf -q$.
(States with $\bf -q$ on the hemisphere of the Fermi surface with
$\bf q \cdot \hat{n}_S >0$, which we may call the ``left'' hemisphere,
are coupled to states of $\bf +q$ on the opposite ``right''
hemisphere due to the
indeterminate sign for the quantum anomaly.)  Thus we may treat each
$({\bf q,~ -q})$ sector separately.  Now we wish to compute the retarded
Green's function:
\begin{equation}
G_{Ret}([u]; {\bf S; q}, t) \equiv
{{\langle \Psi[u] | J({\bf S; q}, t)~ J({\bf S; -q}, 0)
| \Psi[u] \rangle}\over
{\langle \Psi[u] | \Psi[u] \rangle}}~ \theta(t)
\end{equation}
where
\begin{equation}
J({\bf S; q}, t) = \exp[i~ H_c~ t]~ J({\bf S; q})~ \exp[-i~ H_c~ t]
\end{equation}
is the current operator in the Heisenberg picture.
We may choose $\bf \hat{n}_S \cdot q > 0$ so that $J({\bf S; -q}, 0)$
creates a particle-hole pair
at time $t = 0$ while $J({\bf S; q}, t)$ destroys a pair at a later time
$t > 0$.  The crucial step in our calculation is to ignore operator
ordering within the time evolution operators $\exp[\pm i H_c t]$.  This
approximation is exact
so long as the zero mode has macroscopic occupation
since in this case the errors introduced by ignoring operator ordering
are small compared to the total energy.
In other words, we should think of the operator $J({\bf S; -q})$ as removing
just one quantum out of the large number of quanta that make up the
macroscopic zero mode.
Macroscopic occupation corresponds to
$|u({\bf S; q})|^2 >> \Omega~ |{\bf \hat{n}_S \cdot q}|$ which in physical
terms
means that there are a large number of quanta at each point in momentum space
on the Fermi surface.  (Macroscopic occupancy is possible only in the
$\omega$-limit of $\lambda >> |{\bf q}|$; in the opposite q-limit
the Pauli exclusion principle keeps the occupancy small.)
Assuming macroscopic occupancy we have:
\begin{equation}
\exp\{i ~H_c~ t\}~ J({\bf S; q})~ \exp\{-i~ H_c~ t\} =
J({\bf S; p})~ \exp\{i~ E[u({\bf S; q})]~ t\}
\end{equation}
where $E[u]$ is the c-number energy given by
\begin{eqnarray}
E[u({\bf S; q})] &=&
{{\langle \Psi[u] | [J({\bf S; q}, t),~ H_c]~ J({\bf S; -q}, 0)
| \Psi[u] \rangle}\over
{\langle \Psi[u] | J({\bf S; q}, t)~ J({\bf S; -q}, 0) | \Psi[u] \rangle}}
\nonumber \\
&=& {\bf q \cdot \hat{n}_S} \bigg{[} v_F + {{2 \Omega}\over{V}}~
{{1}\over{u({\bf S; q})}}~
\sum_{\bf T} f_c({\bf S - T}) u({\bf T; q}) \bigg{]}\ . \label{energy}
\end{eqnarray}
Thus the Green's function is:
\begin{equation}
G_{Ret}([u]; {\bf S; q}, t) = |u({\bf S; q})|^2~ \exp\{i~ E[u]~ t\}~ \theta(t)
\ ;
\end{equation}
its poles in frequency space
at $\omega = E[u]$ clearly correspond to solutions
of the collective mode equation Eq. (\ref{motc}) in the semiclassical limit.

\section{Quantized Bosons}

In this section we calculate the Green's function in the quiescent state
in the absence of macroscopic
excitations.  In this case the semiclassical
approximation is inapplicable
and the problem must be treated quantum mechanically.
To simplify the calculation we restrict our attention to the
case of spherical (circular in two dimensions) Fermi surfaces and only a single
Fermi liquid parameter, the constant term $F_0$.
Furthermore we consider only
spinless fermions and
again set the Fermi velocity equal to one.
None of these simplifications is essential.

First we write the currents in terms of boson operators that satisfy
canonical commutation relations.  The choice:
\begin{equation}
J({\bf S; q}) = \sqrt{\Omega~ |{\bf \hat{n}_S \cdot q}|}~
[ a({\bf S; q})~ \theta({\bf \hat{n}_S \cdot q}) + a^\dagger({\bf S; -q})~
\theta(-{\bf \hat{n}_S \cdot q}) ] \label{canon}
\end{equation}
with
\begin{equation}
[ a({\bf S; q}),~ a^\dagger({\bf T; p}) ] = \delta^{D-1}_{\bf S, T}~
\delta^D_{\bf q, p} \ ,
\end{equation}
and $\theta(x) = 1$ if $x > 0$ and is zero otherwise,
satisfies the U(1) Kac-Moody commutation relations Eq. (\ref{kacc}) up to
a factor of 2 which does not appear here since the fermions are spinless.
The Hamiltonian Eqs. (\ref{hamc}) and (\ref{intc}) can now be written as
$H_c = H_0 + H_{\rm int}$ where:
\begin{equation}
H_0 = \sum_{\bf q}~ \bigg{\{}
\sum_{\bf S}~ \theta({\bf \hat{n}_S \cdot q})~
({\bf \hat{n}_S \cdot q})~ a^\dagger({\bf S; q}) a({\bf S; q})
+ \sum_{\bf S}~ \theta(-{\bf \hat{n}_S \cdot q})~
(-{\bf \hat{n}_S \cdot q})~ a^\dagger({\bf S; -q})~ a({\bf S; -q})
\bigg{\}}
\end{equation}
and
\begin{eqnarray}
H_{\rm int} &=& \sum_{\bf q}~ \bigg{\{}
g_R~ \sum_{\bf S, T}~ \theta({\bf \hat{n}_S \cdot q})~
\theta({\bf \hat{n}_T \cdot q})~
\sqrt{({\bf \hat{n}_S \cdot q}) ({\bf \hat{n}_T \cdot q})}~
a^\dagger({\bf S; q})~ a({\bf T; q})~
\nonumber \\
&+& g_L~ \sum_{\bf S, T}~ \theta(-{\bf \hat{n}_S \cdot q})~
\theta(-{\bf \hat{n}_T \cdot q})~
\sqrt{(-{\bf \hat{n}_S \cdot q}) (-{\bf \hat{n}_T \cdot q})}~
a^\dagger({\bf S; -q})~ a({\bf T; -q})~
\nonumber \\
&+& g~ \sum_{\bf S, T}~ \theta({\bf \hat{n}_S \cdot q})~
\theta(-{\bf \hat{n}_T \cdot q})~
\sqrt{({\bf \hat{n}_S \cdot q}) (-{\bf \hat{n}_T \cdot q})}~
a({\bf S; q})~ a({\bf T; -q})~
\nonumber \\
&+& \overline{g}~ \sum_{\bf S, T}~ \theta(-{\bf \hat{n}_S \cdot q})~
\theta({\bf \hat{n}_T \cdot q})~
\sqrt{(-{\bf \hat{n}_S \cdot q}) ({\bf \hat{n}_T \cdot q})}~
a^\dagger({\bf S; -q})~ a^\dagger({\bf T; q})~
\bigg{\}}
\end{eqnarray}
with couplings $g_R = g_L = g = \overline{g}
\equiv f_0 {{\Lambda^{D-1}}\over{(2 \pi)^D}}$.  It will be convenient to
denote $a({\bf S; q})$ and $a({\bf S; -q})$ by $a_R({\bf S; q})$ and
$a_L({\bf S; q})$, respectively the right and left moving fields.

The generating functional for the zero temperature correlation functions
is given by an integral over the coherent state eigenvalues
$a_i({\bf S; q}, t)$ and $a^*_i({\bf S; q}, t)$:
\begin{equation}
Z = \int {\cal D}a^*~ {\cal D}a~ \exp \bigg{\{} i \int_{-\infty}^\infty dt~
\big{[} i a_i^* {{\partial}\over{\partial t}}~ a_i - H_c(a^*, a) \big{]}
\bigg{\}}
\end{equation}
where there is an implicit sum over i = L, R and the patch index $\bf S$
which has been suppressed.  The momentum-frequency space propagator
\begin{equation}
i G_i({\bf S; q}, \omega) = \langle a_i({\bf S; q}, \omega)~
a^\dagger_i({\bf S; q},
\omega) \rangle
\end{equation}
is related to the propagator of the $\phi$ fields by:
\begin{equation}
\langle \phi_i({\bf S; q}, \omega)~ \phi_i({\bf S; -q}, -\omega) \rangle
= {{\Omega}\over{4 \pi {\bf \hat{n}_S \cdot q}}}~
\langle a_i({\bf S; q}, \omega)~ a^\dagger_i({\bf S; q}, \omega) \rangle\ .
\label{gphi}
\end{equation}
We now calculate the propagator perturbatively with the use of the bare
right and left propagators:
\begin{eqnarray}
i G_R^0({\bf S; q}, \omega) &=&
\langle a_R({\bf S; q}, \omega)~ a^\dagger_R({\bf S; q}, \omega) \rangle_0
\nonumber \\
&=& {{i}\over{\omega - {\bf \hat{n}_S \cdot q} + i \eta~ {\rm sgn}(\omega)}}
\end{eqnarray}
and
\begin{eqnarray}
i G_L^0({\bf S; q}, \omega) &=&
\langle a_L({\bf S; q}, \omega)~ a^\dagger_L({\bf S; q}, \omega) \rangle_0
\nonumber \\
&=& {{i}\over{\omega + {\bf \hat{n}_S \cdot q} + i \eta~ {\rm sgn}(\omega)}}\ .
\end{eqnarray}
The bare propagators are depicted in Fig.\ \ref{fig1}(i).

At first order in $H_{\rm int}$ there is only one connected contribution to
the right two-point function and it is given by:
\begin{equation}
i G_R^{(1)} = (-i)~ (i G_R^0)~ g_R~ ({\bf \hat{n_S} \cdot q})~ (i G_R^0)
\end{equation}
where we have suppressed the patch, momentum, and frequency labels of
the Green's function.  Amputating the external legs, we find that the first
order contribution to the self energy is just: $\Sigma^{(1)}
({\bf S; q}, \omega)
= g_R~ ({\bf \hat{n}_S \cdot q})$.  At higher orders it is easy to see
that the anomalous couplings $g$ and $\overline{g}$ occur in pairs.
In particular at second order there are two contributions to the one-particle
self energy, and these are shown in Fig.\ \ref{fig1}(ii).
As we build up the complete
set of contributions to the right moving propagator, we split each
scattering process, for example those shown in
Fig.\ \ref{fig2}
into a forward scattered contribution (which involves an intermediate state
in same patch $\bf S$) and a
remainder in which the boson has been scattered into a virtual state in
a different patch.  We can then construct the Dyson
equation, depicted schematically in Fig.\ \ref{fig3},
where the irreducible self
energy $\Sigma^I({\bf S; q}, \omega)$ comprises all amputated diagrams that
cannot be split into two by cutting a single bare right moving propagator.
At the second order the contribution to the irreducible self energy is
therefore:
\begin{equation}
\Sigma^{(2)}({\bf S; q}, \omega) = g^2~ ({\bf \hat{n}_S \cdot q})~ \bigg{\{}
\sum_{\bf T \neq S} \theta({\bf \hat{n}_T \cdot q})~
({\bf \hat{n}_T \cdot q})~ G_R^0({\bf T; q}, \omega)
+ \sum_{\bf T} \theta(-{\bf \hat{n}_T \cdot q})~
(-{\bf \hat{n}_T \cdot q})~ G_L^0({\bf T; q}, \omega)\bigg{\}}\ .
\label{2se}
\end{equation}
Now we specialize to the case of two spatial dimensions. The sums over
patches can be converted to integrals in the $N \rightarrow \infty$ limit
where $\Lambda \rightarrow 0$ as:
\begin{equation}
\Lambda \sum_{S} = k_F \int_0^{2 \pi} d\phi = 2 \pi N(0) \int_0^{2\pi} d\phi
\end{equation}
where $N(0) = {{k_F}\over{2 \pi}}$ for spinless fermions in units where
$v_F = 1$.  The second order contribution to the self energy
can now be written more concisely as:
\begin{equation}
\Sigma^{(2)}(S; {\bf q}, \omega) = -f_0~ {{f_0~ \Lambda}\over{(2 \pi)^2}}~
({\bf \hat{n}_S \cdot q})~ \chi^0({\bf q}, \omega)
\end{equation}
where
\begin{eqnarray}
\chi^0(x) &=& N(0)~ \int_0^{2\pi} {{d\phi}\over{2 \pi}}~
{{\cos(\phi)}\over{\cos(\phi) - x - i \eta~ {\rm sgn}(\omega)}} \nonumber \\
&=& N(0)~ \bigg{\{} 1 - |x| {{\theta(x^2 - 1)}\over{\sqrt{x^2 - 1}}}
+ i |x| {{\theta(1 - x^2)}\over{\sqrt{1 - x^2}}} \bigg{\}}
\nonumber \\
&\equiv& N(0)~ \Omega_0(x)
\label{chi0}
\end{eqnarray}
is the two-dimensional Lindhard function and $x \equiv {{\omega}\over{|{\bf
q}|}}$.
The exact solution of the Dyson equation Fig.\ \ref{fig3} is then given by:
\begin{equation}
\Sigma^I(S; {\bf q}, \omega) = {{f_0~ \Lambda}\over{(2 \pi)^2}}~
({\bf \hat{n}_S \cdot q})~ [1 - f_0~ \chi({\bf q}, \omega)]
\label{self}
\end{equation}
where
\begin{equation}
\chi({\bf q}, \omega) = {{\chi^0({\bf q}, \omega)}\over{1 + f_0~
\chi^0({\bf q}, \omega)}}\ .
\end{equation}
Here we see the equilibrium
Fermi liquid stability criterion $F_0 = f_0 N(0) > -1$ is
necessary to keep the self energy non-singular in the q-limit of $|x| << 1$.
A little algebra then shows that the exact right moving boson propagator can
be written in the compact form:
\begin{equation}
i G_R(S; {\bf q}, \omega) = {{i}\over{\omega - {\bf \hat{n}_S \cdot q}
\bigg{\{} 1 + {{f_0 \Lambda}\over{(2 \pi)^2}}~ [ 1 - f_0~ \chi({\bf q},
\omega) ] \bigg{\}}}}
\end{equation}
Quasiparticle damping occurs in the q-limit
when the Lindhard function has
an imaginary part.  In this regime we may write:
\begin{eqnarray}
{\rm Im}~ f_0 \chi(x) &=& {\rm Im}~ \bigg{\{}
{{F_0 \Omega_0(x)}\over{1 + F_0 \Omega_0(x)}} \bigg{\}}
\nonumber \\
&=&
{\rm Im}~ \bigg{\{} {{A_0 \Omega_0(x)}\over{1 - A_0 [1 - \Omega_0(x)]}}
\bigg{\}}
\nonumber \\
&\approx& A_0 |x|
\end{eqnarray}
where $A_0 \equiv {{F_0}\over{1 + F_0}}$ and the boson
Green's function then reads:
\begin{equation}
i G_R(S; {\bf q}, \omega) = i \bigg{\{} \omega - v_F^\prime~
{\bf \hat{n}_S \cdot q}
+ i {\bf \hat{n}}_S {\bf \cdot q}~ {{A_0^2 \Lambda |\omega|}\over{2 \pi k_F
|{\bf q}|}} \bigg{\}}^{-1}
\end{equation}
where the velocity is slightly renormalized from its bare value of unity:
$v_F^\prime = 1 + F_0 (1 - F_0)~ {{\Lambda}\over{2 \pi k_F}}$.
The boson lifetime is now finite because of
scattering into different
patches.  Note, however, that as the self energy Eq. (\ref{self}) scales to
zero as $\Lambda \rightarrow 0$ it represents an irrelevant correction.
In particular,
the pole in the boson propagator remains unchanged as $N \rightarrow
\infty$.
We expect this to be true generally, regardless of the shape of
the Fermi surface, the details of the Fermi liquid parameters, or whether
the fermions have spin or not.  The renormalization group calculation of
the second section therefore holds, without alteration, when the Fermi
liquid interactions ZS and ZS$^\prime$ are turned on.

\section{Fermion Quasiparticle Properties}

In the previous section we saw that Fermi liquid interactions modify
the boson propagator.  Though large-angle scattering processes were ignored,
small-angle scattering processes made the boson lifetime finite.
With these results we can use
the bosonization formula Eq. (\ref{bosonization}) to infer the fermion
quasi-particle lifetime.  Since bosonization is
carried out in $({\bf x}, t)$ space we must carry out three operations.
First we Fourier transform the boson propagator into real
space.  Next the exponential of the resulting expression yields the
fermion propagator in real space.
Finally an inverse Fourier transform of the fermion propagator back
into momentum space allows us to extract the self energy.

It is difficult technically to perform these steps in all generality.
It will be sufficient for our purposes to first expand the boson propagator
in powers of $f_0$, perform the three operations on each term, and then
reassemble the pieces to find the fermion self energy.  Further, as
we are interested only in the leading (second order) contribution to the
imaginary part of the self energy, we can avoid the first of the two
Fourier transforms.  The
real-space and time boson Green's function may be written as:
\begin{eqnarray}
i G_\phi(S; {\bf x}, t) \equiv
\langle \phi(S; {\bf x}, t)~ \phi(S; {\bf 0}, 0)~
-~ \phi^2(S; {\bf 0}, 0) \rangle
&=& FT~ \big{\{} i G_\phi \big{\}}
\nonumber \\
&=& FT~ \big{\{} i G_\phi^{(0)} + i G_\phi^{(1)} + i G_\phi^{(2)} + ...
\big{\}}
\end{eqnarray}
where $FT$ represents the Fourier transform operation that converts
the variables $({\bf q}, \omega)$ and to $({\bf x}, t)$.
In the second line,
$i G_\phi$ which is given by Eq. (\ref{gphi}), has been expanded
in powers of $f_0$.  The Fourier transform of the leading term,
$FT \big{\{} i G_\phi^{(0)} \big{\}}$, is given by
Eq. (\ref{gphi0}).
Rather than Fourier transforming the first and second order corrections, we
exponentiate this expression to obtain the fermion propagator:
\begin{eqnarray}
i G_\psi(S; {\bf x}, t) &\equiv& \langle \psi^\dagger(S; {\bf x}, t)~
\psi(S; {\bf 0}, 0) \rangle \nonumber \\
&=& {{\Omega}\over{V~ a}}~ e^{i {k_S\cdot x}}~
\exp \big{\{} {{4 \pi}\over{\Omega^2}}~ i G_\phi(S; {\bf x}, t) \big{\}}
\nonumber \\
&=&  {{i \Lambda}\over{(2 \pi)^2}}~
{{e^{i k_F x_\parallel}}\over{{\bf x \cdot \hat{n}}_S
- t + i \delta~ {\rm sgn}(t)}}~ \exp \bigg{\{}
{{4\pi}\over{\Omega^2}}~
FT~ \big{[} i G_\phi^{(1)} + i G_\phi^{(2)} + ... \big{]} \bigg{\}}
\nonumber \\
&=&  {{i \Lambda}\over{(2 \pi)^2}}~
{{e^{i k_F x_\parallel}}\over{{\bf x \cdot \hat{n}}_S
- t + i \delta~ {\rm sgn}(t)}}~
\bigg{\{}1 +
i FT~ [\tilde{G}_\phi^{(1)}] - {{1}\over{2}}~ (FT~ [\tilde{G}_\phi^{(1)}])^2
+ i FT~ [\tilde{G}_\phi^{(2)}]
+ O(f_0^3) \bigg{\}}
\label{ft}
\end{eqnarray}
where in the last two lines we have assumed $|x_\perp \Lambda| << 1$ and
in the last line we have absorbed the factor of $4\pi/\Omega^2$ into
$\tilde{G}_\phi^{(i)} \equiv (4\pi/\Omega^2) G_\phi^{(i)}$.
We are interested primarily in the imaginary part of the fermion
self energy.  The first order contribution
to the boson self energy contained in $i G_\phi^{(1)}$ is purely real
and therefore does not contribute.  The leading contribution to the
imaginary part of the fermion self energy comes from $i G_\phi^{(2)}$
which is given by:
\begin{equation}
i G_\phi^{(2)}(S; {\bf q}, \omega) = -i {{\Omega}\over{4\pi}}~
{{f_0^2~ \Lambda}\over{(2\pi)^2}}~ \chi^0({\bf q}, \omega)~
[\omega - q_\parallel + i \eta~ sgn(\omega)]^{-2}\ .
\end{equation}
Since $|x_\perp \Lambda| << 1$ this
contribution to the boson propagator must be integrated over $q_\perp$,
using Eq. (\ref{chi0}) for $\chi^0$ and we obtain:
\begin{equation}
i~ {\rm Im}~ \int_{-\Lambda/2}^{\Lambda/2} dq_\perp~
\chi^0(S; q_\perp, q_\parallel, \omega)
= -i N(0)~ |\omega|~ \ln |{{\omega^2 - q_\parallel^2}\over{\Lambda^2}}|\ .
\end{equation}
The appearance of the logarithm in this equation is peculiar to
two spatial dimensions.  In three dimensions the integral is a
two-dimensional one over the two coordinates perpendicular to the
Fermi surface normal and the
imaginary part of the Fermion self energy is proportional simply to
$\omega^2$.

We now take the inverse Fourier transform $FT^{-1}$.
Expanding the fermion propagator as
$G_\psi \equiv G^0_\psi + \delta G_\psi + ...$, the leading
imaginary contribution to the fermion propagator in
$(k_\parallel, \omega)$ space is given by:
\begin{eqnarray}
i \delta G_\psi(k_\parallel, \omega) &=& i{{f_0^2 \Lambda
N(0)}\over{(2\pi)^3}}~
\int_{-\infty}^{\infty} d\omega^\prime~
\int_{-\infty}^{\infty} dq_\parallel~ |\omega^\prime|~
\ln |{{\omega^{\prime 2} - q_\parallel^2}\over{\Lambda^2}}|
\nonumber \\
&\times& {{1}\over{(\omega^\prime - q_\parallel + i \eta~ {\rm sgn}(
\omega^\prime))^2~ [(\omega^\prime - \omega) - (q_\parallel - k_\parallel)
+ i \eta~ {\rm sgn}(\omega^\prime - \omega)]}}\ .
\end{eqnarray}
The integral may be performed by complex integration; no divergences occur
since all the poles in the complex $q_\parallel$ plane lie either on
one side of the real axis or the other unless $\omega^\prime$ lies
between $0$ and $\omega$.
Thus, except for this limited range of frequencies, the contour in the
$q_\parallel$ plane may be closed without enclosing any poles.  The result
is:
\begin{eqnarray}
i \delta G_\psi(k_\parallel, \omega) &=&
-{{1}\over{2}}~ {{f_0^2 N(0)}\over{(2\pi)^2}}~
{{1}\over{[\omega - k_\parallel + i \eta~ {\rm sgn}(\omega)]^2}}~
{\rm sgn}(\omega)~ \bigg{\{}
[\omega^2 + (\omega - k_\parallel)^2/4]~ \ln {{|\omega - k_\parallel|}
\over{\Lambda}}
\nonumber \\
&+& [\omega^2 - (\omega - k_\parallel)^2/4]~ \ln {{|\omega + k_\parallel|}
\over{\Lambda}}
-{{\omega}\over{2}}~ (2 \omega - k_\parallel)
\bigg{\}}
\end{eqnarray}
and therefore the fermion self energy at this order is given by:
\begin{eqnarray}
{\rm Im}~ \Sigma^{(2)}_f(k_\parallel, \omega) &=&
{{1}\over{2}}~ {{f_0^2 N(0)}\over{(2 \pi)^2}}~
{\rm sgn}(\omega)~ \bigg{\{}
[\omega^2 + (\omega - k_\parallel)^2/4]~ \ln {{|\omega - k_\parallel|}
\over{\Lambda}}
+ [\omega^2 - (\omega - k_\parallel)^2/4]~ \ln {{|\omega + k_\parallel|}
\over{\Lambda}}
\nonumber \\
&-& {{\omega}\over{2}}~ (2 \omega - k_\parallel)
\bigg{\}}\ .
\label{fself}
\end{eqnarray}
The imaginary part of the self energy at the quasiparticle pole is the
inverse of the quasiparticle lifetime.
The location of the pole has been shifted from its bare value to
$\omega = v_F^\prime k_\parallel$ due to renormalization of the Fermi
velocity, $v_F^\prime = 1 + F_0 (1 - F_0)~ {{\Lambda}\over{2 \pi k_F}}$.
As a result the imaginary part of the self energy at the pole is given by:
\begin{equation}
{\rm Im}~ \Sigma^{(2)}_f(\omega)|_{\rm pole} =
{{1}\over{2}}~ {{f_0^2 N(0)}\over{(2\pi)^2}}~
{\rm sgn}(\omega)~ \bigg{\{}
\omega^2~ \ln {{F_0~ \omega^2}\over{\pi \Lambda k_F}} - {{\omega^2}\over{2}}
\bigg{\}}
\end{equation}
the form of which we immediately recognize from
previous work on two-dimensional Fermi liquids\cite{hodges,bloom,Stamp2}.
This quantity is always negative since $\omega^2 << \Lambda k_F$.
Despite the appearance of the logarithm, the weight of the
quasiparticle pole, $Z_k$, remains non-zero at the Fermi surface;
the regular Fermi liquid interaction $F_0$ does not destroy the Fermi liquid
fixed point.  We expect that more general regular Fermi liquid
interactions, the inclusion of the spin sector, or
extensions to non-spherical Fermi surfaces will
not change this result qualitatively.

\section{Singular Interactions}

Singular Fermi liquid
interactions in 2D were proposed by Anderson\cite{PWA} and studied
perturbatively by Stamp\cite{Stamp1,Stamp2}.  The interaction studied
couples opposite spins and
diverges as $\bf k \rightarrow k^\prime$:
\begin{equation}
f^{\sigma \sigma^\prime}({\bf k k^\prime})
= {b\over N(0)}\delta _{\sigma ,-\sigma ^\prime }
{({\bf k} + {\bf k^\prime}) \cdot ({\bf k-k^\prime})\over{|{\bf k-k^\prime}|^2
}}~ \theta (|{\bf k}|-k_F) ~\theta (k_F-|{\bf k^\prime}|) \label{singul} \ .
\end{equation}
Note that both $\bf k$ and $\bf k^\prime$ lie off the Fermi surface
(respectively above and below it).  Thus this interaction is of a more
general sort than the type Landau originally envisaged in the phenomenological
theory.  The interaction diverges like ${1\over{|{\bf k
- k^\prime}|}}$ as ${\bf k} \rightarrow {\bf k^\prime}$.
It is related to the {\it regular} interaction that arises at second order
in a Taylor-series expansion of the Landau function.  In three dimensions
this regular interaction gives rise to
a $T^3 \ln T$ contribution to the specific heat\cite{Pethick,Tony}:
\begin{eqnarray}
f({\bf k, k^\prime}) &=& a + b~ ({\bf \hat{p} \cdot \hat{q}})^2 + ...
\nonumber \\
&=& a + b~ \bigg{\lbrace}
{({\bf k} + {\bf k^\prime}) \cdot ({\bf k} - {\bf k^\prime})\over
|{\bf k + k^\prime}| |{\bf k - k^\prime}|} \bigg{\rbrace}^2 + ... \label{regul}
\end{eqnarray}
for ${\bf q} \equiv {\bf k - k^\prime}$
and ${\bf p} \equiv {\bf k + k^\prime}$.
This interaction vanishes if ${\bf k}$ and ${\bf k^\prime}$ lie
on the Fermi surface and approach the same point.  If, on the other hand,
both momenta lie away from the surface then the interaction approaches a
non-zero, but finite, limiting value as the two momenta converge.

The bosonized Hamiltonians Eq. (\ref{hamc}) and Eq. (\ref{hams})
generalize Fermi liquid theory in a
different way: $\bf S$ and $\bf T$
lie on the Fermi surface but $\bf q$ need not be zero.  Nevertheless, the
interactions mentioned above have natural counterparts in the bosonized
theory.  The $T^3 \ln T$
contribution to the specific heat is recovered in this picture\cite{Tony}
by setting
${\bf k} = {\bf k_S + q}$ and ${\bf k^\prime } = {\bf k_T - q}$
in Eq. (\ref{regul})
to obtain:
\begin{equation}
V_c({\bf S, T; q}) = {{1}\over{2}}~ \Omega^{-1}~ v_F~ \delta^{D-1}_{\bf S,T}
+ {{1}\over{V~ N(0)}}~ \bigg{\{} a + 4b~ {({\bf \hat{n}_S \cdot q})^2\over
({\bf k_S - k_T})^2 + 4 q^2} + ... \bigg{\}}\ .
\end{equation}
Note that this contribution to the specific heat is a sub-leading
correction which just reflects the fact noted above
that any $\bf q$-dependence of regular Landau parameters is irrelevant
to the leading-order behavior.
That both generalizations yield the same non-analytic thermodynamic
behavior suggests that
they are equivalent up to irrelevant terms.  Therefore we proceed to make
the same substitution in the singular interaction Eq. (\ref{singul})
of Anderson and Stamp to find in the spin sector:
\begin{equation}
V_s({\bf S, T; q}) = {{1}\over{2}}~ \Omega^{-1}~ v_F~ \delta^{D-1}_{\bf S,T}
+ {{b}\over{V~ N(0)}}~ {k_F^2~ |2~ {\bf \hat{n}_S \cdot q}~ /~ k_F|^\beta\over
({\bf k_S - k_T})^2 + 4 q^2}\ \label{sings}
\end{equation}
where $\beta$ interpolates between the Anderson-Stamp interaction ($\beta = 1$)
and the regular interaction ($\beta = 2$) (now for the spin sector).
We will assume that $S \neq T$ in the second term; otherwise the
linear dispersion relation is destroyed at the outset and
consider the scaling of the largest part of the interaction as we
take $N \rightarrow \infty$.  Observing that the largest contribution
comes from nearest-neighbor patches $\bf S$ and $\bf T$, and
$|{\bf q}| \leq \lambda << \Lambda$, we see that the interaction scales as
\begin{equation}
{{b}\over{V~ N(0)}}~ N^{2 \alpha - \beta}\ ,
\end{equation}
where the exponent $\alpha$ was defined in section II by the equation
$\Lambda = k_F/N^\alpha$,
and therefore the interaction is singular only if $\beta < 2 \alpha < 2$.

The Anderson-Stamp interaction is singular only as $S \rightarrow T$.
An example of an interaction which is singular for all $S$ and $T$ and which
has more obvious physical significance is afforded by the Coulomb interaction.
The bare interaction may be factorized into contributions to the three
channels (ZS, ZS$^\prime$, and BCS).  We assume
that the BCS channel renormalizes to zero since it is repulsive.
Furthermore, for small $|{\bf q}|$ the ZS$^\prime$ exchange channel is
much smaller than the ZS direct channel.  In this limit we find:
\begin{equation}
V_c({\bf S, T; q}) = {{1}\over{2}}~ \Omega^{-1}~ v_F~ \delta^{2}_{\bf S,T}
+ {{1}\over{V}}~ {{e^2}\over{4 \pi |{\bf q}|^2}} \label{coulomb}
\end{equation}
in three spatial dimensions with $V_s$ containing only regular interactions.
Of course the identification of the bare
Coulomb interaction with the small-angle scattering amplitude neglects the
physics of screening.
Nevertheless it would be interesting to determine the
effect of the bare interaction Eq. (\ref{coulomb})
on the single quasiparticle lifetime.
If the technical problem of performing
the Fourier transforms mentioned in the previous section can be overcome,
the bosonization method would yield non-perturbative insight into the effect
of singular interactions on the Fermi liquid.

\section{Discussion}

The Coulomb interaction Eq. (\ref{coulomb})
mentioned in the previous section illustrates the
difference between collective modes and
single-particle excitations.  If we neglect screening,
naively substitute the Coulomb interaction
Eq. (\ref{coulomb}) into the charge collective mode Eq. (\ref{motc}), and
compute the spectrum, we find in three dimensions
a gap comparable to the plasma frequency; see, for example\cite{Rick}.
Thus charged Fermi liquids do not support low energy
collective modes in the charge sector.  Nevertheless,
we know that the single-particle spectrum remains gapless.  Thus, the
spectrum of single-particle bosonic excitations about the quiescent state,
unlike the collective modes, must also remain gapless.

The fact that the bosonized Hamiltonian separates into a sum of charge and
spin parts, $H = H_c + H_s$,
raises the specter that, as in one dimension\cite{Tony},
the quasiparticle propagator might also
exhibit spin-charge separation, even in the case of regular Fermi liquid
interactions.  Spin-charge separation in dimensions larger than
one would, however, destroy the Green's function approach to
Fermi liquid theory as the key element in that approach, the existence of a
pole in the single-particle Green's function with spectral weight $0 < Z < 1$,
would be replaced by a branch cut and $Z$ would equal zero.
Fortunately this does not happen because, as we saw at the end of
section IV, the location of the
pole of the boson propagator is unchanged from its free value
in the $\Lambda \rightarrow 0$ limit.  Consequently the spin and
charge velocities are equal and spin charge separation does not occur.

Finally we note that
our bosonic analysis of the renormalization group flows near the
Fermi liquid fixed point does not rely on a particular form for the fermion
propagator.  This is in contrast with Shankar's approach which assumes that
the one-particle propagator always retains a Fermi liquid form; consequently
non-Fermi liquid fixed points are ruled out from the outset of the calculation.
The bosonized theory
contains non-perturbative information; only technical difficulties
prevent us from evaluating directly the non-perturbative fermion
propagator.  It should be possible to surmount these difficulties.

\section{Acknowledgements}
J.B.M. thanks Andreas Ludwig, Andrei Ruckenstein, and Phil Nelson for helpful
discussions at the Aspen Center for Physics and H.-J.K. thanks Doochul Kim for
hosting his stay at Seoul National University.
This research was supported by the National Science Foundation
through grant DMR-9008239 (A.H.) and DMR-9357613 (J.B.M.).

\begin{figure}
\caption{Boson Green's functions.  (i) Right and left moving bare boson
propagators $G^0_R({\bf S; q}, \omega)$ and $G^0_L({\bf S; q}, \omega)$.
(ii) The two second order contributions to the self energy
which involve virtual states on the
the right and left sides of the Fermi surface.}
\label{fig1}
\end{figure}

\begin{figure}
\caption{Self energy at second and third order for the boson propagator.
(i) The second order contribution to the self energy
which involves virtual states on the right side of the Fermi surface.
The first diagram on the right hand side of the equation
(with two crosses) represents scattering
into and out of the same patch $S$.
The second diagram represents scattering into and out of
a {\it different} patch $T \neq S$ (denoted by a dashed line with a slash).
(ii) Some of the third order contributions to the self energy.  Not shown
are contributions which involve virtual states on the left side of the Fermi
surface.  Of the diagrams shown, only the first (with two dotted lines)
contributes to the irreducible self energy $\Sigma^I$; the remaining three
diagrams break into two pieces when one of the bare propagators is cut.}
\label{fig2}
\end{figure}

\begin{figure}
\caption{The Dyson equation for the self energy.  The double line represents
the exact one-particle boson propagator.}
\label{fig3}
\end{figure}


\begin{references}
\bibitem{PWAbook}P. W. Anderson, {\bf Basic Notions of Condensed Matter
Physics}, p. 84, Benjamin-Cummings Publishing Co. (London, 1984).
\bibitem{Haldane}F. D. M. Haldane, unpublished.
\bibitem{Tony}A. Houghton and J. B. Marston, {\it Phys. Rev. B} {\bf 48},
7790 (1993).
\bibitem{Shankar}R. Shankar, {\it Physica} A{\bf 177}, 530 (1991) and
to appear in {\it Rev. Mod. Phys.}
\bibitem{Others}G. Benfatto and G. Gallavotti, {\it Phys. Rev. B} {\bf 42},
9967 (1990); J. Feldman and E. Trubowitz, {\it Helv. Phys. Acta} {\bf 63},
157 (1990); ibid, {\bf 64}, 213 (1991); ibid, {\bf 65}, 679 (1992).
\bibitem{Luther}A. Luther, {\it Phys. Rev. B}, {\bf 19}, 320 (1979).
\bibitem{Fradkin}A. H. Castro Neto and Eduardo Fradkin, ``Bosonization of
Fermi Liquids,'' Univ. of Illinois preprint, 1993.
\bibitem{Baym}Gordon Baym and Christopher Pethick, {\bf Landau Fermi Liquid
Theory: Concepts and Applications}, John Wiley \& Sons (New York, 1991).
\bibitem{sugawara}H. Sugawara, {\it Phys. Rev.} {\bf 170}, 1659 (1968).
\bibitem{Alex}A. E. Meyerovich and K. A. Musaelian, {\it J. Low Temp. Phys.}
{\bf 89}, 781 (1992).
\bibitem{hodges}C. Hodges {\it et al.}, {\it Phys. Rev. B} {\bf 4}, 302
(1971).
\bibitem{bloom}P. Bloom, {\it Phys. Rev. B} {\bf 12}, 125 (1975).
\bibitem{Stamp2}P. C. E. Stamp, {\it J. Phys. I} {\bf 3}, (1993).
\bibitem{PWA}P. W. Anderson, {\it Phys. Rev. Lett.} {\bf 64}, 2306 (1990)
and {\it Phys. Rev. Lett.} {\bf 66}, 3226 (1991).
\bibitem{Stamp1}P. C. E. Stamp, {\it Phys. Rev. Lett.} {\bf 68}, 2180 (1992).
\bibitem{Pethick}C. J. Pethick and G. M. Carneiro, {\it Phys. Rev. A} {\bf 7},
304 (1973).
\bibitem{Rick}G. Rickayzen, {\bf Green's Functions and Condensed Matter},
pp. 163 -- 167, Academic Press Limited (London, 1980).
\end{references}
\end{document}